\documentclass{article}
\usepackage{e}
\usepackage{graphics}
\begin{document}
\branch{A}
%
\title{Pictures of quantum nuclear rotation
beyond the correspondence principle}
\author{R. Arvieu\inst{1}\and P.~Rozmej\inst{2,3}}
\institute{Institut des Sciences Nucl\'eaires, 53 av. des Martyrs,  
 F--38026 Grenoble-Cedex, France\\
e-mail: arvieu@in2p3.fr
\and Theoretical Physics Department,  University Maria 
 Curie--Sk\l odowska, PL--20031 Lub\-lin,  Poland\\
e-mail: rozmej@tytan.umcs.lublin.pl
\and Gesellschaft f\"ur Schwerionenforschung mbH,  
 Planckstr.\ 1, D--64291 Darmstadt, Germany\\ 
e-mail: p.rozmej@gsi.de}
\maketitle
 \begin{abstract}
We analyze the time evolution of simple nuclear rotational wave packets
(WP) called circular, linear or elliptic, depending on
squeezing parameter $\eta$, assuming that $E=\hbar\omega_0\, I(I+1)$.
The scenario of fractional revivals found by Averbukh and Perelman is 
adapted to symmetric WP and compared to that which holds for asymmetric 
WP. In both cases various shapes are identified under these lines in 
particular many cases of cloning. 'Mutants' WP are found most often.
Finally the time evolution of a WP formed by Coulomb excitation
on $^{238}$U and calculated by semiclassical theory is also presented.
\end{abstract}

\section{Introduction}\label{intro}

	It is somewhat paradoxical after many years of confirmation of
quantum mechanics that new features can be learned on two of the simplest
systems that are taught in the elementary courses: the hydrogen atom and
the rigid rotor. It was indeed realized only in the early 90's
\cite{dacic,boris}
that a wave packet in the hydrogen atom evolves in time according to a
universal scenario discovered theoretically by Averbukh and Perelman
\cite{averbukh}. For
short times a well built wave packet moves around a Kepler ellipsis, 
and then it spreads along the ellipsis. 
The surprise was that the wave packet is
seen at specific times as a superposition of fractional wave packets
concentrated around specific points on this trajectory. This behaviour
results from the unitary quantum evolution and from self interferences.
By changing the initial wave packet one can also obtain a radial wave packet
without an analogue classical trajectory which will also experience the
scenario of fractional revivals \cite{yeazell}.
An extension of these results has
later been been made by Bluhm and Kostelecky \cite{bluhm} who showed
the existence  of superrevivals for longer times.

	The aim of this article is to describe the similar steps for the
rotation of symmetric rigid rotors with possible application to nuclei 
and to molecules.	
The theory of ref.~\cite{averbukh} is exact
for all times if the energy spectrum is quadratic in one quantum
like the infinite square well \cite{aronstein} or in two quantum
numbers like \cite{agarwal} or the rigid top with axial symmetry. 
In ref.~\cite{rozar} we studied various sets of angular momentum
coherent states which represent an heteronuclear diatomic molecule or a
rigid body with axial symmetry.
Our purpose is to show the rotation of a symmetric object
described by a quantum pure state in the most obvious scheme (i.e. if the
law $E=(\hbar^2/2J)(I(I+1)$) holds exactly) for a wave packet specifically 
designed to rotate and what the motion is, 
if the initial preparation of the wave packet is different.
Because of the
quantum spreading a quantum WP does not rotate at all like a classical
rigid body, its spreading differs according to the initial conditions of
its preparation. The introduction of this spreading is necessary if one
introduces the concept of a WP which is not a common practice in the field
of rotational nuclei or molecules. We must however acknowledge that this
concept was first introduced in the past in the reference book by
Broglia and Winther~\cite{broglia} 
on heavy ions collisions, then in a more
extensive work by Fonda et. al.~\cite{fonda} 
who have discussed the mechanism of
generation of several sets of rotational coherent WP. Our work completes
these two works in the sense that we give a full study of the time
evolution of the WP which is now possible after the work of 
\cite{averbukh}. 
Our WP depend also on a new squeezing parameter $\eta$ 
introduced in \cite{rozar} 
which enables to change the 
geometry of the WP in a most significant manner.

  Before proceding to our work we must also quote most elegant papers by
Berry and Golberg \cite{berrygold} 
and Berry \cite{berry1} 
who have studied the full time
evolution of the propagator of a nuclear spin with a hamiltonian of the
form $L^2/(2J)$. More recently various Talbot effects were also discovered
with the help of renormalization techniques \cite{berryklein}. 
The occurrence of fractal
dimensions both in space and time during the evolution of uniform WP in
boxes of arbitrary dimensions has been also discovered \cite{berry2}. 
These last results show that the spreading of WP in a bound system 
is extremely rich. Our purpose is to emphasize this richness 
for two dimensional WP on the sphere. 
  
\section{Asymmetric and symmetric squeezed WP}\label{sec:2}

\subsection{Asymmetric WP}\label{sec:2.1}
In ref.~\cite{rozar} we have studied  the properties and the time
evolution of the following WP called the exponential coherent WP which
depends on two parameters $N$ and $\eta$.
This normalized state is defined as
\begin{equation} \label{b6} 
\Psi_{\eta}(\theta, \phi) = \sqrt{\frac{2N}{4\pi \sinh (2N)} }\,
 {\mathrm{e}}^{N\sin\theta(\cos\phi+{\mathrm{i}}\eta\sin\phi) } \; .
\end{equation}
For large enough values of $N$ the probability density associated to it is
symmetric around $Ox$ and is peaked within a solid angle
$\Omega=4\pi/(4N+1)$.
Thus, large values of $N$ correspond to defining the
orientation of the axis of symmetry of the system in a very sharp
way, quite independently on the value of the average angular momentum 
$\vec{L}$. For the components of L one can see that
\begin{equation} \label{b2} 
  \langle L_x \rangle =0 \hspace{10ex} \langle L_y \rangle =0 \; ,
\end{equation}
\begin{equation} \label{b3} 
  \langle L_z \rangle = \eta (N-\frac{1}{2})	 \; .
\end{equation}
 Equation (\ref{b3}) is correct only for large $N$, 
 for small $N$ there is a small
correction term not important for our discussion. 
Therefore, the average of $\vec{L}$ is fixed by the product $\eta N$. 
Finally we have proved \cite{rozar} 
that the state (\ref{b6}) is made of a special class of angular 
momentum states called in the literature by the name of intelligent 
spin states (for real values of $\eta$). They obey the equation
\begin{equation} \label{b5} 
(L_x+{\mathrm{i}}\eta L_y)\Psi_{\eta +} = 0 \; .
\end{equation}
From this equation one can derive that the parameter $\eta$ has the meaning
of a squeezing parameter since 
\begin{equation} \label{n5} 
            |\eta|^2=\frac{\Delta L_x^2}{\Delta L_y^2}  
\end{equation}
For all real values of $\eta$ the minimum uncertainty condition 
is satisfied
\begin{equation} \label{b4} 
\Delta L_x^2 \:\Delta L_y^2= \frac{1}{4}\,
\langle L_z \rangle ^2 \;.
\end{equation}
The states (\ref{b6}) are such the analogous of the squeezed states of the
electromagnetic field.
In this article we will distinguish two limiting values of $\eta$:
the case $\eta=0$ corresponds to a purely real WP initially
without momentum which we call the linear WP, 
the case $\eta=1$, called a circular WP, is an eigenstate of $L_+$. 
The other values correspond in the same terminology
\cite{rozar} to elliptic WP. Indeed these WP can be derived from Gaussian
WP which are respectively associated to linear, circular and elliptic  
trajectories if they evolve in 3D harmonic oscillator.

  The case of complex values of $\eta$ does not bring any interesting 
freedom \cite{arro}, 
the main difference lying in the fact that the equality (\ref{b4}) is
violated. Therefore $\eta$ will take only real values in the following. 

 The spherical harmonic expansion of the state (1) is written 
in terms of coefficients
$b_{IM}$ as
\begin{equation} \label{b7} 
\Psi_{\eta} = \sum_{IM} \, b_{IM} \, Y^I_M(\theta,\phi) \; ,
\end{equation}
\begin{eqnarray} \label{b8} 
b_{IM} & = & \sqrt{\frac{2N}{\sinh (2N)}}  \\
& & \times\, \sum_{ll'} \,(-1)^{l'} \,
N^{l+l'} \, \frac{(1+\eta)^l (1-\eta)^{l'}}{\sqrt{(2l)!(2l')!}} \,
\frac{\langle ll'00|I0\rangle \langle ll'l-l'|IM\rangle}{\sqrt{(2I+1)}}
\; . \nonumber
\end{eqnarray}
For $\eta=1$ only the $b$'s with $M=I$ are non-zero. This is why these wave
packets are called circular. In the case $\eta=0$ it is simpler to make a
rotation of the coordinate axis to get an eigenstate of $L_z$ with
eigenvalue 0, the expansion of $\Psi_0$ takes the form
\begin{equation} \label{b9} 
\Psi_0 = \sum_{I} \, b_{I0} \, Y^I_0(\theta,\phi) \; .
\end{equation}
with $b_{I0}$ given in terms of spherical 
Bessel function of the first kind by
\begin{equation} \label{b10} 
b_{I0}  =  \sqrt{\frac{2N}{\sinh (2N)}} \, \sqrt{(2I+1)} \,
\sqrt{\frac{\pi}{2N}}\, I_{I+1/2}(N)   \; .
\end{equation}
 
\subsection{Symmetric WP}\label{sec:2.2}
The situation we want to describe is that of an even,
deformed nucleus with an axis of symmetry $O_z$ such that 
$\langle L_Z\rangle=K=0$ and with a
plane of symmetry perpendicular to this axis. No other degree of freedom
will be included. We therefore consider instead of (\ref{b6}) the following
symmetrical combination 
\begin{equation} \label{b1} 
\Psi_{\eta +}(\theta, \phi) = \frac{C_{+}}{2}\left 
[{\mathrm{e}}^{N\sin\theta(
\cos\phi+{\mathrm{i}}\eta\sin\phi)} +
{\mathrm{e}}^{-N\sin\theta(\cos\phi+
{\mathrm{i}}\eta\sin\phi)}\right ] \; .
\end{equation}
Clearly for large $N$ it represents two antipodal waves, within the same
solid angle as above and with the same symmetry around $Ox$. 
The normalization
constant $C_+$ depends now both on $N$ and $\eta N$ because of an 
interference term between the two waves. It is given by
\begin{equation} \label{n12} 
 C_+= \left \{ 2\pi \left [\frac{\sinh 2N}{2N}+\frac{\sin 2\eta N}{2\eta N}
 \right ] \right \}^{(-1/2)}  
\end{equation}
It should be stressed that the main difference in the time evolution of
(\ref{b6}) and (\ref{b1}) comes from the interference between these 
two parts. As we will see it will become much different from a 
normalization factor.
 
  It is straightforward to verify that eqs. 
(\ref{b2}) to (\ref{b4}) are still valid.
 We now express the wave packet $\Psi_{\eta +}$
 in terms of new coefficients $b_{IM+}$ as
\begin{equation} \label{b11} 
\Psi_{\eta +} = \sum_{IM} \, b_{IM+} \, Y^I_M(\theta,\phi) \; ,
\end{equation}
 and these coefficients are related to the $b_{IM}$ by
\begin{equation} \label{b12} 
b_{IM+} = \left\{ \begin{array}{ll}
 0 & \mbox{~~for $I$ odd} \\
 \sqrt{2} \,\left (1+\frac{\sin 2\eta N}{\eta\sinh 2N}\right )^{-1}\,
  b_{IM} \simeq \sqrt{2} \, b_{IM}
 & \mbox{~~for $I$ even.} \end{array} \right.
\end{equation}

\section{Time evolution}\label{sec:3}

\subsection{The rigid rotor assumption}\label{sec:3.1}
  Let us study the evolution of $\Psi_{\eta +}$ for $\eta=0$
and $\eta=1$
assuming that the energy levels obey the rule
\begin{equation} \label{b13} 
E_I=\frac{\hbar^2}{2J} \,I(I+1) \; .
\end{equation}
 In the following we will use $\omega_0=\hbar/(2J)$ and we define an average
angular momentum by
\begin{equation} \label{b14} 
\bar{I}(\bar{I}+1) = \langle \Psi_{\eta +}|L^2| \Psi_{\eta +} \rangle \; .
\end{equation}
We can now use the work of \cite{averbukh} to analyze
the structure of the wave
packets at time $t$ knowing simply that the wave packets are given by
\begin{equation} \label{b15} 
 \Psi_{\eta +}(\theta,\phi,t)= \sum_{IM}\, b_{IM +}\, Y^I_M(\theta,\phi)
\, {\mathrm{e}}^{{\mathrm{-i}}\omega_0\, I(I+1)t}
\end{equation}
It is necessary to introduce two characteristic times as in \cite{averbukh}
\begin{equation} \label{b16} 
T_{{\mathrm{cl}}} = \frac{2\pi}{\omega_0}\, \frac{1}{2\bar{I}+1} ,\hspace{10ex}
T_{{\mathrm{rev}}} = \frac{2\pi}{\omega_0}\; .
\end{equation}
Note that $T_{{\mathrm{rev}}}>>T_{{\mathrm{cl}}}$ if $\bar{I}$ 
is large enough.
We have with (\ref{b15}) and (\ref{b16})
\begin{equation} \label{b17} 
 \Psi_{\eta +}(\theta,\phi;T_{{\mathrm{rev}}})= 
 \Psi_{\eta +}(\theta,\phi;0)\; .
\end{equation}

\subsection{The outline of the theory of fractional revivals}\label{sec:3.2}
 Let us sketch the general theory of \cite{averbukh} which allows
to find a universal scenario of fractional revivals.  
 Near $t=(m/n)T_{\mathrm{rev}}$, where $m/n$ is an irreducible ratio of two
integers, it has  been proved in \cite{averbukh} that
\begin{equation} \label{b18} 
{\mathrm{e}}^{{\mathrm{-2i}}\pi I^2\,\frac{m}{n}} = \sum_{s=0}^{l-1}\: a_s \:
{\mathrm{e}}^{-2 {\mathrm{i}} \pi I\frac{s}{l}}
\end{equation}
This sum contains $q=\frac{n}{2}$ non-zero terms if $n$ is even
and $q=n$ terms if $n$ is odd. The integer $l=\frac{n}{2}$ if $n$  
is a multiple of 4, $l=n$ in the other cases (therefore the difference
between $l$ and $q$ is necessary to take into account that the $a_s$
with even $s$ are zero if $n$ is even and not multiple of 4).
The modulus of $a_s$ is simply for all the cases where it is non-zero
equal to:
\begin{equation} \label{b19} 
|a_s| = \frac{1}{\sqrt{q}}
\end{equation}
The phase of $a_s$ as a function of $m,n$ and $s$ can be found in an
older paper written in a quite different context \cite{hannay}.
Therefore one writes after the insertion of (\ref{b18}) into (\ref{b7})
taken at time $t=\frac{m}{n} \, T_{\mathrm{rev}}$ 
\begin{eqnarray} \label{b20} 
\Psi_{\eta}(\theta,\phi;\frac{m}{n} T_{\mathrm{rev}}) & = & 
\sum_{s=0}^{l-1} \: a_s \,\left [ \sum_{IM} \, b_{IM} \, Y^I_M \, 
{\mathrm{e}}^{{\mathrm{-2i}}\pi 
I(\frac{m}{n}+\frac{s}{l})} \right ] \\
& = & \sum_{s=0}^{l-1} \: a_s \: \Psi^{(s)}_{\mathrm{cl}}(\theta,\phi;t_s) \;.
\label{b21} 
\end{eqnarray}
Any wave packet is a sum of $q$ fractional wave packets
 $\Psi^{(s)}_{\mathrm{cl}}$,
each with a different effective time $t_s$
\begin{equation} \label{b22} 
t_s = (\frac{m}{n}+\frac{s}{l})\, T_{\mathrm{rev}} \; .
\end{equation}
The fractional wave packet is defined as 
\begin{equation} \label{b23} 
\Psi_{\mathrm{cl}}^{(s)}(\theta,\phi;t_s) = \sum_{IM} \, b_{IM} \, Y^I_M \, 
{\mathrm{e}}^{{\mathrm{-i}}\omega_0 It_s}\; .
\end{equation}

\subsection{Clones and mutants}\label{sec:3.3}
There are several cases for which $\Psi_{\mathrm{cl}}$ 
is a clone of the initial
$\Psi$ defined by (\ref{b6}) or (\ref{b7}):
\begin{description}
\item[i)] 
if the expansion (\ref{b7}) contains only $b_{IM}$ with the single
values $M=I$, this is the case for the wave packet (\ref{b6}) for $\eta=1$.
One obtains for each $t_s$
\begin{equation} \label{b24} 
\Psi_{\mathrm{cl}}(\theta,\phi;t_s) = 
\Psi_{\eta}(\theta,\phi-\omega_0 t_s)\; .
\end{equation}
This result is independent of the distribution of the $b_{II}$'s,
it applies to all kinds of circular wave packets. The sum (\ref{b21})
is a superposition of $q$ clones of the initial wave packet. In
\cite{aronstein} this conclusion was derived for an infinite 
one-dimensional well. In our case with $\eta=1$ the $q$ clones are
localised around $q$ axis forming a symmetric fan in the $Oxy$ plane.
\item[ii)] 
if $\omega_0 t_s$ is a multiple of $2\pi$, i.e. if $a_{s_0}$ is non-zero
for the value
\begin{equation} \label{b25} 
s_0 = n - m \; ,
\end{equation}
then
\begin{equation} \label{b26} 
\Psi_{\mathrm{cl}}(\theta,\phi;t_{s_0}) = \Psi_{\eta}(\theta,\phi,0) \; .
\end{equation}
This occurs when $n$ is odd or even and not multiple of 4. For these times
there is a clone at the place of the initial wave packet with the 
amplitude $a_{s_0}$. The existence of this clone is independent of the 
$b_{IM}$ and the parameter $\eta$.
\end{description}

In \cite{rozar} we have proposed to call mutants the fractional wave packets
$\Psi_{\mathrm{cl}}$ which are not clones and which have a different geometry
due to the phase ${\mathrm{e}}^{{\mathrm{-i}}\omega_0 It_s}$ in (\ref{b23}). We have also
described the change of the shape of mutants as a function of $\eta$
for fixed $t_s$. This is made by considering two extreme cases where the shape
of $\Psi_{\mathrm{cl}}$ is simply understood. 
For $\eta=1$ $\Psi_{\mathrm{cl}}$ is a clone
of the initial wave packet located around a given direction in the $Oxy$
plane. For  $\eta=0$ the expansion (\ref{b9}) shows that wave packet
$\Psi_0$ at $t=0$ as well as for all $t$ and also each 
$\Psi_{\mathrm{cl}}$
have cylindrical symmetry around $Ox$. $\Psi_{\mathrm{cl}}$ is therefore 
peaked around one or two rings around $Ox$. For intermediate values of $\eta$  
the mutants have shapes making a smooth transition between the two extreme
cases. 

This theory fits perfectly to the case of the asymmetric WP which
contains terms with even and odd values of $I$ in equations  (\ref{b7}),
(\ref{b18}) and (\ref{b20}). In the case $\frac{m}{n}=\frac{1}{2}$
eq.\ (\ref{b22}) gives $t_s=T_{\mathrm{rev}}$,
 $a_s=\delta_{s\,1}$ and the
WP is periodic with period $t_s=T_{\mathrm{rev}}/2$. 
This reduced periodicity
can be seen immediately from the fact that $I(I+1)$ is always even.
It is therefore enough to consider times such that $\frac{m}{n}\leq
\frac{1}{2}$.

\subsection{Additional rules for symmetric WP}\label{sec:3.4}
In the symmetric case corresponding to (\ref{b1}) and (\ref{b11}) with
(\ref{b12}), an additional symmetry arises from the fact that eq.\ (\ref{b18}) 
is now used for even values of $I$. If we put $n=4n'$ in (\ref{b18})
we obtain $q=l=2n'$ values of $s$ in the general case. However if $I$
is always even $l$ takes the value of $n'/2$ if $n'$ is a multiple of 4
or $n'$ is odd or even and not multiple of 4. For $n'$ even and not multiple of
4 the $a_s$ with even $s$ are zero. Therefore the number $q$ of fractional 
waves is largely reduced and becomes 
\begin{equation} \label{b27} 
q = \left\{ \begin{array}{ll}
 n' & \hspace{5ex}\mbox{for odd $n'$} \\
 n'/2 & \hspace{5ex}\mbox{for even $n'$.} \end{array} \right.
\end{equation}
For $n'=1$ one can write 
\begin{equation} \label{b28} 
{\mathrm{e}}^{{\mathrm{-2i}}\pi \, I(I+1)/4} = 
{\mathrm{e}}^{{\mathrm{i}}\pi \, I/2} \; ,
\end{equation}
for $n'=2$
\begin{equation} \label{b29} 
{\mathrm{e}}^{{\mathrm{-2i}}\pi \, I(I+1)/8} = 
{\mathrm{e}}^{{\mathrm{-i}}3\pi \, I/4} \; .
\end{equation}
Inserting these values in (\ref{b15}) for the particular case of circular WP
($\eta=1$) one obtains the following periodic behaviour
\begin{equation} \label{b30} 
\Psi_{1+}(\theta,\phi,T_{\mathrm{rev}}/4) = \Psi_{1+}(\theta,\phi+\pi/2,0) \; ,
\end{equation}
\begin{equation} \label{b31} 
\Psi_{1+}(\theta,\phi,T_{\mathrm{rev}}/8)= \Psi_{1+}(\theta,\phi-3\pi/4,0) \; .
\end{equation}
For $\eta\neq 1$ the insertion of the phases (\ref{b27})
into (\ref{b15}) produces a {\em single} WP different from the initial one,
i.e.\ a single mutant. For $\eta=1$ the time interval necessary to study
the evolution is $t=T_{\mathrm{rev}}/8$, but for $\eta\neq 1$ it is still
$t=T_{\mathrm{rev}}/2$.

At fractional times of $T_{\mathrm{rev}}/4$, like 
$t=T_{\mathrm{rev}}\,\frac{m}{4n'}$ 
the following formula holds for $\eta=1$ and describes fully the cloning
of these species of symmetric WP
\begin{equation} \label{b32} 
\Psi_{1+}(\theta,\phi,T_{\mathrm{rev}}\,\frac{m}{4n'})= \sum_{s=0}^l \, a_s \,
\Psi_{1+}(\theta,\phi-\pi(\frac{s}{l}+\frac{m}{2n'}),0) \; ,
\end{equation}
with the general result that the sum contains $q=n'/2$ non-zero terms
for even $n'$ (with $l=n'/2$ if $n'$ is a multiple of 4, $l=n'$ in other
cases and $a_s$ is zero if $n'$ is even and not multiple of 4 and $s$ is even).

For example the application of this formula for  $t=T_{\mathrm{rev}}/16$ gives
two clones oriented along perpendicular directions in the $Oxy$ plane
since 
\begin{equation} \label{b33} 
\Psi_{1+}(\theta,\phi,T_{\mathrm{rev}}/16)=  a_0 \,
\Psi_{1}(\theta,\phi-\pi/8,0) + a_1 \, \Psi_{1}(\theta,\phi-5\pi/8,0)\; .
\end{equation}
Similarly, for $t=T_{\mathrm{rev}}/24$ one obtains 3 clones etc...
For $\eta\neq 1$ the clones should be replaced by a corresponding number of
mutants.

If on the other hand $n$ is even and not multiple of 4, i.e.\ if
$n=2n'$ with $n'$ odd, the limitation of even $I$ in eq.\ (\ref{b18}) leads
to $l=n'$ therefore all the $a_s$ are non-zero but should be calculated with 
an effective ratio $2m/n'$ instead of $m/n$. The number of fractional WP is 
then the same as for the asymmetric case. Similarly, if $n$ is odd there are 
$q=n$ fractional WP. Note that for odd $n$ and even $n$ not multiple of 4
eq.\ (\ref{b25}) and (\ref{b26}) also hold: one of the fractional WP 
is a clone at the place of the initial WP.

\section{Numerical calculations for the exponential coherent WP.}
\label{sec:4}

\subsection{Comparisons between the symmetric and asymmetric WP}
\label{sec:4.1}
  As already stated above, the WP (\ref{b1}) differs from (\ref{b6}) by the
interference terms which are almost negligible if one calculates the
probability density at t=0. However the previous discussion on the time
evolution shows a large number of differences. Therefore a
comparison between asymmetric and symmetric WP is presented in the
figures of this letter. With high enough $N$ one obtains a generic 
behaviour. For $\eta=1$ we have chosen $N=14$ i.e.\ $\bar{I}=13.5$.
This value of $\bar{I}$ corresponds to the average excitation
energy of the $K=0$ rotational band of $^{238}$U that is obtained
from Coulomb excitation by 200 MeV $^{40}$Ar projectiles.
Coulomb excitation does not produce weights $b_{IM}$ so simply
distributed to give a WP like (\ref{b1}). 

\begin{figure}
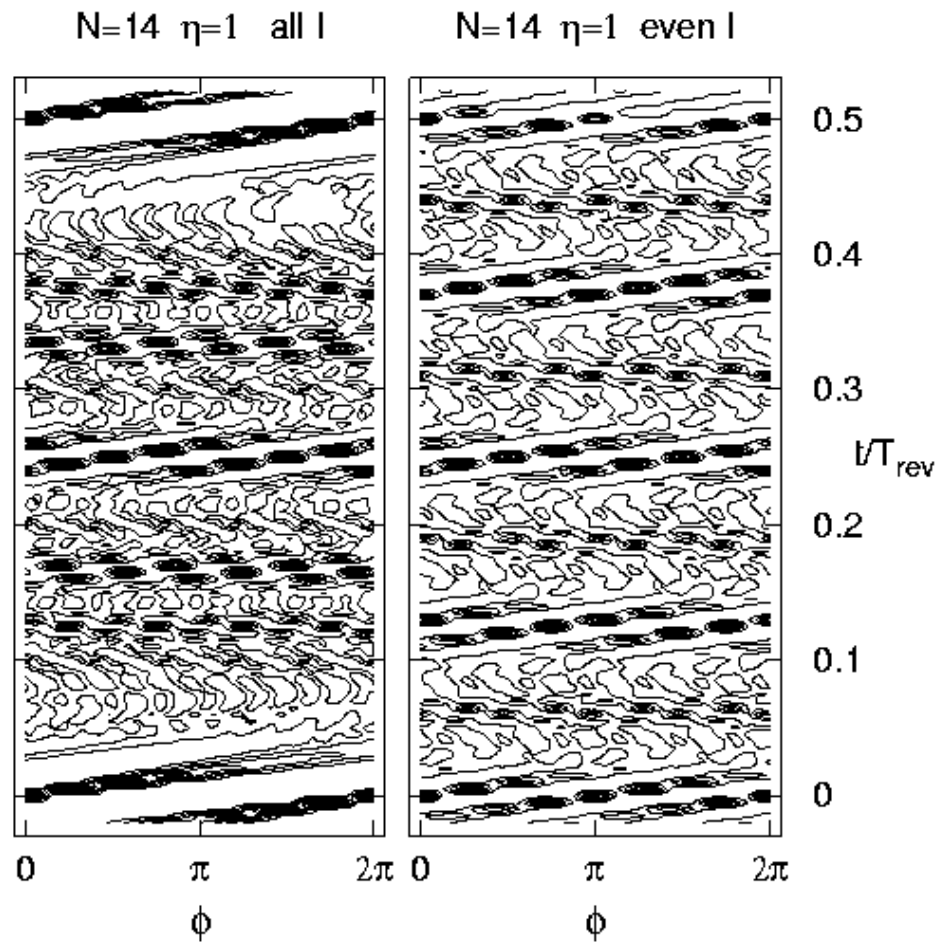

\caption[f1]{Time evolution of circular, asymmetric (left)
and symmetric (right) WP with $N=14$ 
in the 'carpet' representation. 
The $|\Psi(\theta,\phi,t)|^2$ is presented in the contour
plot for fixed $\theta=\pi/2$. }
\label{n14eta1}
\end{figure}

Fig.~\ref{n14eta1} 
shows 'carpets' for the two cases with $N=14$ and 
$\eta=1$. The carpet is the intersection of the plane $Oxy$
with the probability density as a function of time for 
$0<t<T_{\mathrm{rev}}/2$. For 
$t<T_{\mathrm{rev}}/25=T_{\mathrm{cl}}$ the WP spread and 
occupy all values of phi.
For $t>T_{\mathrm{cl}}$ there are clear windows near values 
$(m/n)T_{\mathrm{rev}}$ for which the WP is formed of $q$ 
clones, with $q$ given as discussed in the text for asymmetric 
and symmetric case respectively.
In the former case one has large windows for 
$m/n=1/4$ $(q=2)$, $m/n=1/6$ and $1/3$ $(q=3)$, $m/n=1/8$ and $3/8$
$(q=4)$ and it is possible to see as much as $q=8$ clones 
without much interference between them (see Fig.~\ref{n14eta1asclo}).
In the latter case we obtain instead due to eqs.\ (\ref{b29}) and 
(\ref{b31}) a slide angle of $-3\pi/4$ since the following equation 
takes place
\begin{equation} \label{b34} 
\Psi_{1+}(\theta,\phi,t+T_{\mathrm{rev}}/8) =
\Psi_{1+}(\theta,\phi-3\pi/4,t) \; .
\end{equation}
 Therefore the interval $T_{\mathrm{rev}}/2$ is separated into 
four stripes into which essentially only a single large time window 
is seen for $m/n=1/16$ with $q=n'/2=2$ clones. By zooming within 
$T_{\mathrm{rev}}/8$ one can see other windows as well as we will 
see in Fig.~\ref{n14eta1syclo}.


    Fig.~\ref{n14eta1asclo} and  \ref{n14eta1syclo}
 show the probability density as a function 
of $\theta$ and $\phi$ for a selection of times where one 
can underline similarities or differences between the two cases.
For $0<t<T_{\mathrm{cl}}$ the WP essentially spreads 
in $\phi$ because the corresponding values of $n$ produce too 
many clones with big interferences \cite{aronstein}.
For $t=T_{\mathrm{rev}}/50$ the two peaks forming the symmetric WP
interfere strongly and the WP occupies already all the values of $\phi$.
In the next sequences of times one identifies $(q,q')$ clones respectively 
in the asymmetric and symmetric cases:
$m/n=1/16$ $(8,2)$, $m/n=1/3$ $(3,3)$, $m/n=1/24$ $(12,3)$:
there the 12 clones are strongly interfering,
$m/n=1/5$ $(5,5)$, $m/n=1/6$ $(3,3)$.

 Fig.~\ref{n110eta0} shows the carpets corresponding to 
 $\eta=0$ with $N=110$ and not
 $N=14$. The distribution of the $b_{IM}$ with $I$ depends indeed
 very much on the value of $\eta$. Using $N=14$ with $\eta=0$
 produces too low value of $\bar{I}$ and a rather poor carpet.
 The value $N=110$ corresponds to $\bar{I}=10$ and leads to richer
 figures. For this value of $\eta$ the WP is axially symmetric for all 
 times, the carpet is therefore defined as the quantity 
$2\pi\sin\theta\,|\psi_0(\theta,t)|^2$ as a 
function of $\theta$ and $t$. Both carpets are also symmetric around 
$t=T_{\mathrm{rev}}/4$. However only the carpet showing 
the symmetric case has an additional symmetry around $\theta=\pi/2$.
Indeed a time $\tau$ which fulfills for all I the equation 
\begin{equation} \label{b35} 
 \exp[-{\mathrm{i}}\omega_0\,\tau\,I(I+1)]=(-1)^I  
\end{equation} 
does not exist. The maxima of the $\Psi_{\mathrm{cl}}^{(s)}(\theta,t_s)$ occur 
for definite values of $\theta$ and are
not always distributed symmetrically around $\pi/2$. These fractional waves
can be understood as waves on a sphere with maximum along a ring 
\cite{rozar}. For $t=T_{\mathrm{rev}}/4$ there is a unique large 
ring at $\theta=\pi/2$. Moreover, since $b_{I0}$ and
$Y_{I0}$ are both real, values $s$ and $s'$ may exist such that
\begin{equation} \label{b36} 
       \Psi_{\mathrm{cl}}^{(s')}={\Psi_{\mathrm{cl}}^{(s)}}^*  \; ,
\end{equation} 
	 for example for $t=1/10\,T_{\mathrm{rev}}$ there are 5
fractional WP with $0<s<4$. 
For $s_0=n-m=4$ there is a clone such that $\Psi_{\mathrm{cl}}^{(4)}$
is equal to the initial $\Psi$.
This WP is represented as a dark dot at $\theta=0$
in the left part of Fig.~\ref{n110eta0} 
and there are two pairs of fractional WP, one for $s=0$ and
$s'=4$, the other for $s=2$ and $s'=3$. 
By addition of these 4 WP one obtains only
two rings represented as two other large dots for different $\theta$.
In the symmetric case this number is doubled for $T_{\mathrm{rev}}/10$.
As discussed above the existence of the clone relies upon valus of $m/n$.
 
In the right part of Fig.~\ref{n110eta0} one can most often 
detect a ring at $\theta=\pi/2$. 
Similar carpets have been discussed in \cite{grossmann}
for the infinite well in one dimension.

\subsection{Geometrical shapes of clones and mutants}\label{sec:4.2}
 We present in Fig.~\ref{n14t18} a short selection of shapes of WP for 4 different
values of $\eta$ and the same time $t=T_{\mathrm{rev}}/8$. 
Only the part of the $|\Psi|^2$ 
with $z>0$ is shown, note that $\phi$ is defined as the angle with the $Oy$ axis. 
In the asymmetric case one obtains 4 clones for $\eta=1$. When $\eta$ is 
decreased these clones become 'mutants' which develop a larger spread in 
the angle $\theta$. For $\eta=0$, as well as for $\eta=1/25$,
one obtains two rings 
around the symmetry axis $Oy$ as well as a maximum along this axis. In the 
symmetric case we illustrate for the circular WP the law given by 
eqs. (\ref{b31},\ref{b34}):
there is a periodicity with a slide angle $-3\pi/4$. For different $\eta$ the 
clone is transformed into a single mutant with a larger spread in $\theta$.
For the case of the linear WP this mutant is cylindrically 
symmetric around $Oy$.
Note that the mutant has, for $\eta=1/25$, an interesting dissymmetrical shape
which is necessary in order to make the transition between the clone for 
$\eta=1$ and the two rings for $\eta=0$.

   The WP studied in this letter exibits well defined shapes and a well
understood evolution. This is due to the assumptions that the law (\ref{b13})
holds exactly and that the $b_{IM}$ are simply distributed according 
to (\ref{b8}) in
order to correspond to the simple exponentials (\ref{b1}) or (\ref{b6}).

\section{Calculations with realistic WP}\label{sec:5}

 It is quite possible to improve our study in order to incorporate
ingredients which make it more realistic.
It is well known \cite{broglia,fonda} that during Coulomb excitation (CE)
a deformed nucleus is excited to a coherent mixture of rotational states.
This superposition is also peaked around a mean value of angular momentum,
so one can expect similar features as predicted by scenario of Averbukh and
Perelman. The most clear case is CE with backscattering as in this case 
excited WP has cylindrical symmetry (only $Y^I_0$ components, 
i.e. $\eta=0$ or {\em linear} WP). 
The partial waves and fractional revivals have then
topology of rings on a sphere. Therefore for presentation of shapes only
one angular variable ($\theta$) is sufficient. In Fig. \ref{u238}
we present the time evolution of WP obtained by CE of $^{238}$U bombarded by
$^{40}$Ar at E=170 MeV. 
The amplitudes of excitation of given $I$ angular momentum
eigenstates have been calculated within semiclassical theory of 
CE \cite{broglia}.
The left part shows the `ideal case', i.e. when $E_I$'s follow perfect rotor
dependence $I(I+1)$, the right corresponds to time evolution  
with energies taken from experiment.  
In the latter case it is necessary to define time scales of the
evolution.
In general \cite{averbukh}  
\begin{equation} \label{b37}
T_{\mathrm{rev}} = \frac{2\pi}{\frac{1}{2} | E_{I}''|_{I=\bar{I}}}, \quad 
T_{\mathrm{cl}}=\frac{2\pi}{|E_{i}'|_{I=\bar{I}}} \;. 
\end{equation}
For perfect rotor case these times are determined unambigously:
$T_{\mathrm{rev}}=(2\pi/\hbar^2)J$, $T_{\mathrm{cl}} = T_{\mathrm{rev}}/(2\bar{I}+1)$.
For experimental energies we adopt the following estimation for $T_{\mathrm{rev}}$.
The energies of the ground state rotational band for $^{238}$U are well
approximated by the two parameter formula
\begin{equation} \label{b38}
E_I \simeq  a\,I(I+1) + b\,[I(I+1)]^2     
\end{equation}
with $a= 7.5$ MeV, $b=-0.004$ MeV \cite{nucdat}. 
Our 'estimated' $T_{\mathrm{rev}}$, used for the figures (there is no 
exact revival at all), is then defined by (\ref{b37}) and (\ref{b38})
taken for $\bar{I}$ determined from the equation
\begin{equation} \label{b39}
\bar{I}(\bar{I}+1) = \langle \vec{L}^2\rangle  
\end{equation}
i.e determined by coefficients of CE rotational WP (see Fig. 8, p. 82 of
ref. \cite{broglia}). 
In Fig. \ref{u238short} we display shapes of WP (precisely the probability 
density as a function of $\theta$ angle) for short times, 
showing spreading of the WP, initially localized at poles, and clear few
rings structure of WP at particular times (3/400 and 4/400*$T_{\mathrm{rev}}$).
In Fig. \ref{u238long} we show the same WP after much longer evolution, exhibiting
structures very similar to those observed for very short times. 

Although the 'carpet' for experimental energies isn't as regular as
that for `ideal' ones, still strong revivals of WP occur. The absolute
time scales for nuclear rotation are very short ($T_{\mathrm{rev}}\sim 10^{-19}$s, 
$T_{\mathrm{cl}}\sim 10^{-20}$s) and are still beyond time resolution of present
experimental techniques. It is clear that the
picture which emerges from this study underlines that the time evolution of
nuclear WP depends strongly on their excitation mechanism as well as from
the assumption of a rigid body value for the moment of inertia. The
inclusion of more realistic ingredients introduces less symmetry and less
regularity in the time evolution. 

\section{Conclusions}\label{sec:6}

 There is no doubt that the time evolutions described in
our article have been calculated by observing strictly the rules of
quantum mechanics for a pure state. It is reasonable to believe from the
works of \cite{broglia} and \cite{fonda}
that a WP formed by Coulomb excitation should move as
described in section \ref{sec:5}.
However there is not much hope that the tools of
experimental nuclear physics can evolve in such a way that this time
evolution is observable since one needs to probe the WP after its
formation with the help of a second projectile.
In future the creation of nuclear rotational wave packets and
observation of their evolution should be in principle possible with gamma
lasers.
Nevertheless the picture of the rotation one should have in mind must 
be that given above, i.e. a
picture where the interference effects play a dominant role and differ
according to the symmetry of the rotor and its initial preparation. 

 The field of molecular physics looks more appropriate for observation and
many theoretical calculations as well as several experiments have already
been realized. Rotational coherence effects of large molecules have been
studied theoretically and observed for about ten years 
\cite{felker}--\cite{felkerbas} but are
generally understood as thermal average of rotational quantum beats. The
rotation of a single molecule of O$_2$ \cite{stipe} 
or HB-DC 
\cite{gimzewski} within a supra
molecular bearing has been recently observed. The rotation is induced
thermally and there is no reason to believe that the system is in a pure
state. In the frame of optimal control theory \cite{warren,shen}, 
one has also shown
recently how to create rotational coherent WP of asymmetric--top molecules
by designing tailored microwave pulses \cite{ortigoso,persico}. 
The order of magnitude for the time scales of molecular rotations
varies from $10^{-12} s$ \cite{persico} to $10^{-14} s$ \cite{ortigoso}
and is available for experiment. However, the known excitation mechanisms
lead mainly to rovibrational states and an excited superposition is
usually composed of smaller number of components than rotational
coherent states discussed in this paper.  
A most important
challenge for our purpose would be to find the excitation mechanism that
enables the generation of our squeezed WP. Excitation mechanism of a rigid
rotor with intense laser field is still under study but like Coulomb
excitation of nuclei, this mechanism leads only to linear waves (with
$\eta=0$).

  The study of the time evolution of WP in simple sytems produces
behaviour which is full of surprises. 
A recent work on WP transmitted through a
potential barrier has detected the presence of high momentum components
which are present during a short time \cite{brouard} 
an event which is also a challenge for detection. 
 

{\sf Acknowledgements}
One of us (P. R.) acknowledges financial
support from Centre National de la Recherche Scientifique
during his stay at ISN. 

This work has been supported by KBN Grant No. 2P03B14314.

\vspace{30mm}
{\large\sf High resolution figures (*.tar.gzip) can be obtained as 
MIME attachments
by e-mail from rozmej$@$tytan.umcs.lublin.pl }


\vfill\pagebreak

\begin{figure}
\caption[f2]{Shapes of circular, asymmetric WP in the initial stages of the
evolution and some fractional revival times. Note different
vertical scales used in order to show details. }
\label{n14eta1asclo}
\end{figure}

\begin{figure}
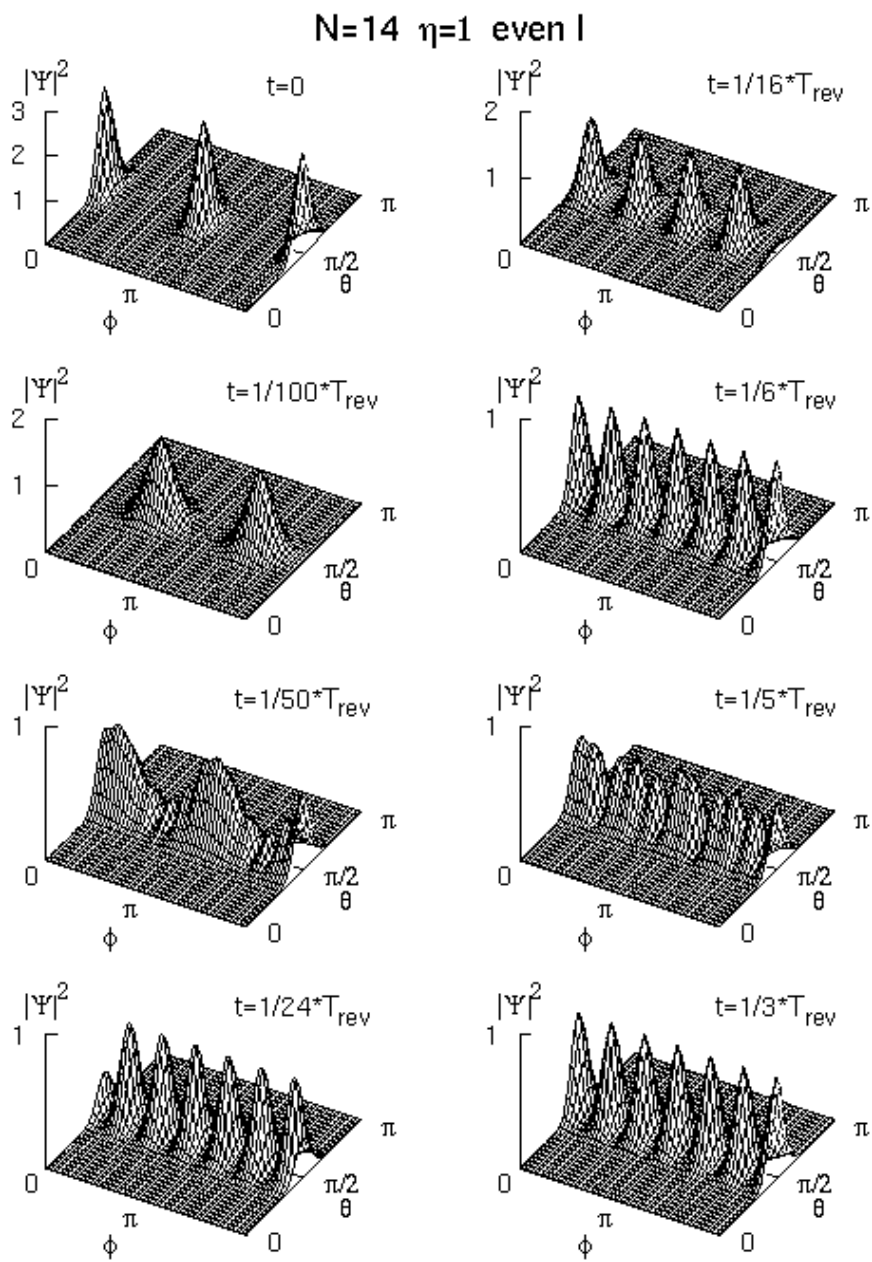

\caption[f3]{ The same as in Fig. \ref{n14eta1asclo} but for symmetric WP.}
\label{n14eta1syclo}
\end{figure}

\begin{figure}
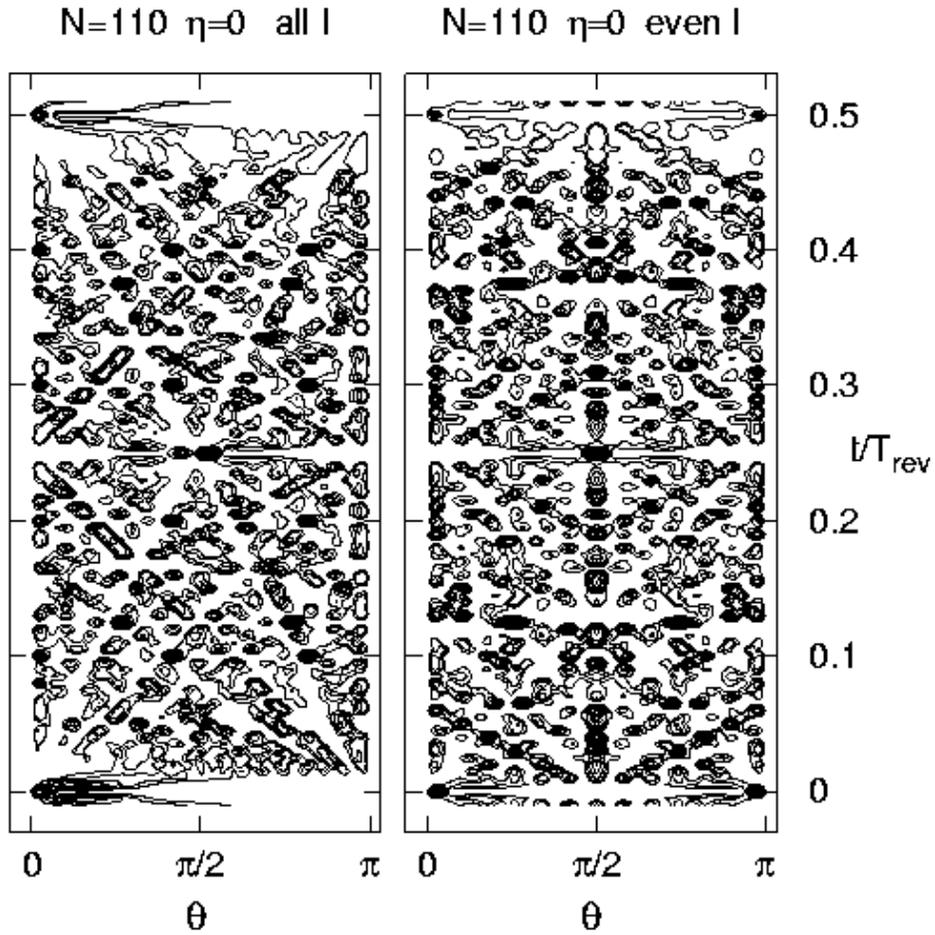

\caption[f4]{
'Carpet' representation of the time evolution of linear, asymmetric
(left) and symmetric (right) WP with $N=110$ and $\eta=0$.
In this case the probability density
 $2\pi\sin\theta |\Psi(\theta,t)|^2$ is presented in the contour
plot. }
\label{n110eta0}
\end{figure}

\begin{figure}
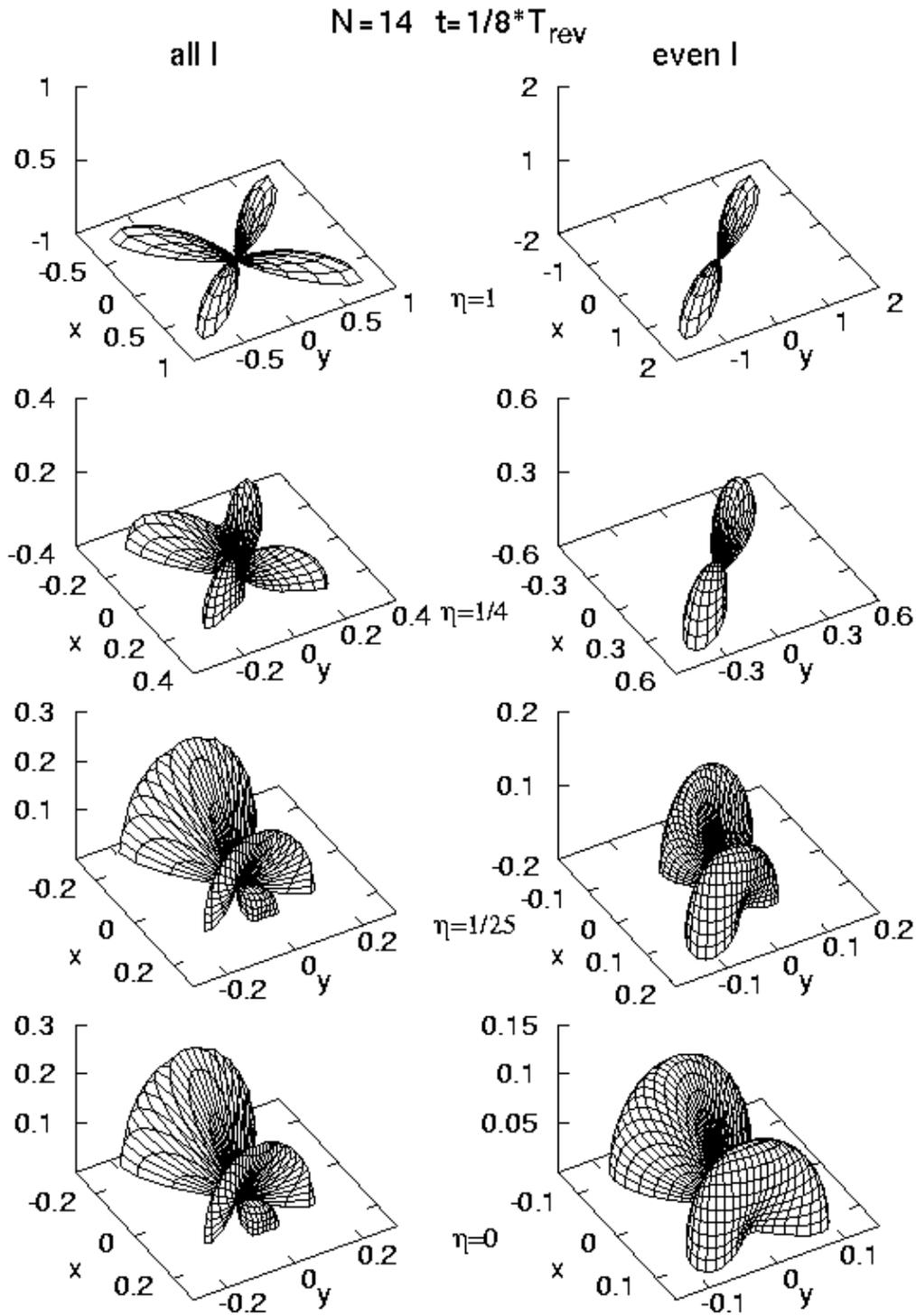

\caption[f5]{Shapes of clones and mutants for asymmetric (left)
and symmetric (right) WP with  $N=14$ and different $\eta$ at
time $t= T_{\mathrm{rev}}/8$. The $|\Psi(\theta,\phi,t)|^2$ is
presented in spherical coordinates as the radial coordinate,
$\theta$ is an angle with respect to $Ox$ axis which is the
axis of cylindrical symmetry in this case. (Only upper part of
the probability density is displayed)}
\label{n14t18}
\end{figure}

\begin{figure}
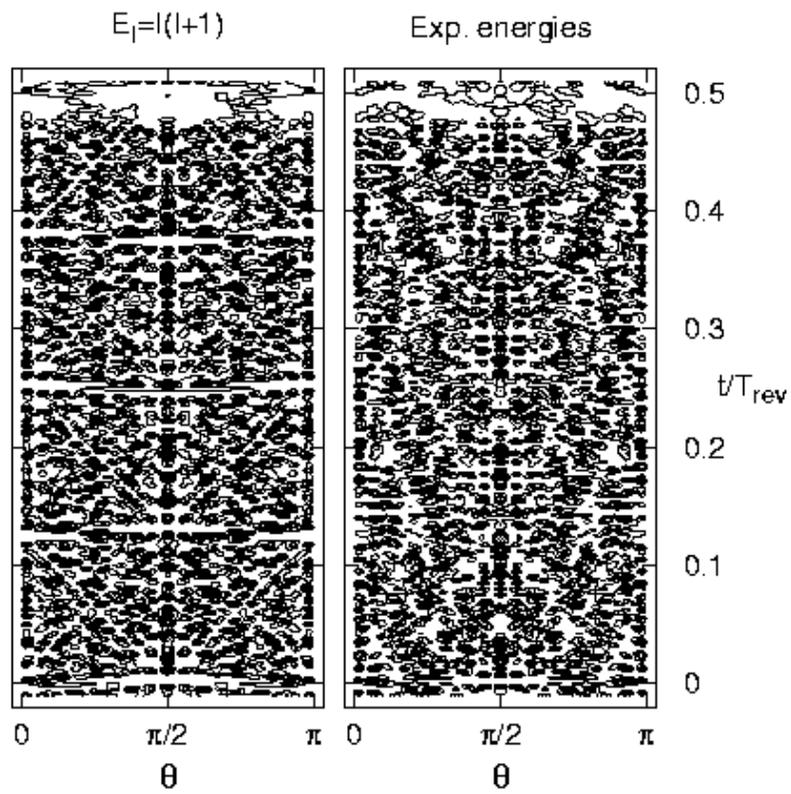

\caption[f6]{Time evolution of nulear rotational wave packet obtained in CE of
$^{238}$U presented in `quantum carpet' representation. Contours of
$2\pi\sin\theta |\Psi|^2$ are plotted.}
\label{u238}
\end{figure}

\begin{figure}
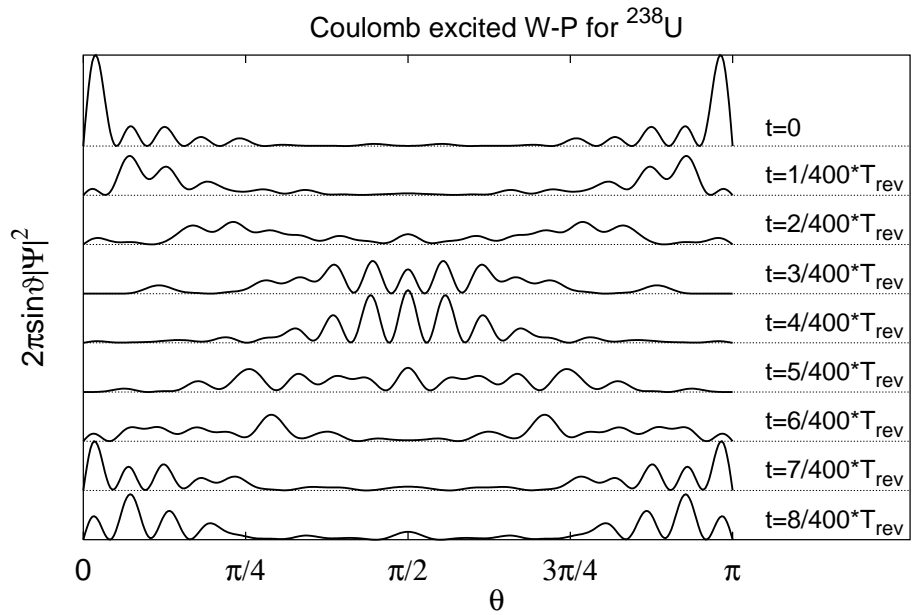

\caption[f7]{Shapes  of nulear rotational wave packet 
obtained in CE of $^{238}$U during a short time evolution.}
\label{u238short}
\end{figure}
\vfill\pagebreak

\begin{figure}
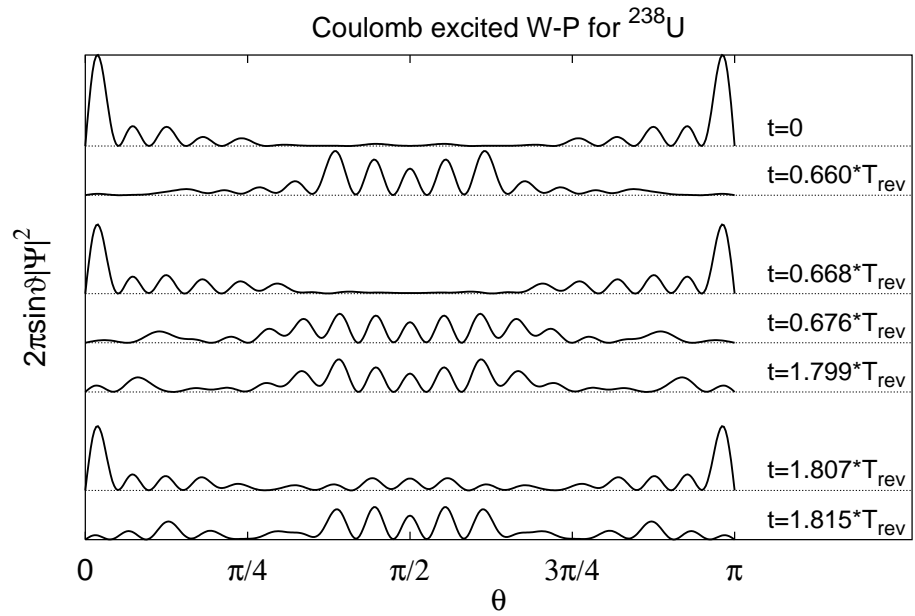

\caption[f8]{The same as in Fig. \ref{u238short} for some particular
much longer times.}
\label{u238long}
\end{figure}

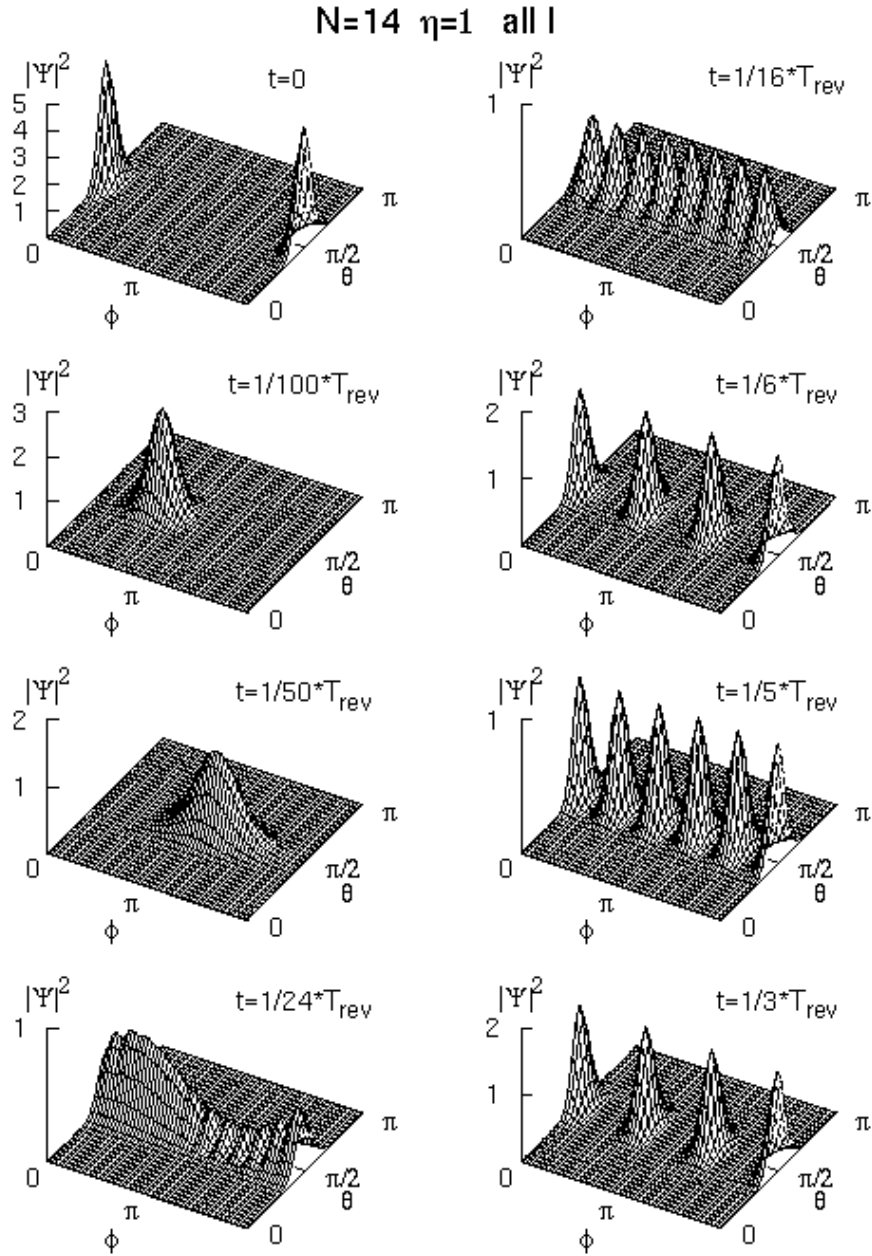
\end{document}